\documentclass[longnamesfirst,apjl]{emulateapj}
\usepackage{apjfonts,natbib,epsfig}
\usepackage{graphics}
\tighten

\usepackage{color}

\newcommand{\beq}{\begin{equation}}
\newcommand{\eeq}{\end{equation}}
\newcommand{\beqa}{\begin{eqnarray}}
\newcommand{\eeqa}{\end{eqnarray}}

\begin{document}

\title{Lensing and Supernovae: Quantifying The Bias on the Dark Energy
Equation of State}  \author{Devdeep Sarkar$^1$, Alexandre Amblard$^1$,
Daniel E.  Holz$^{2,3}$, and Asantha Cooray$^1$} \affil{$^1$Department
of Physics and Astronomy, University of California, Irvine, CA 92617\\
$^2$Theoretical Division, Los  Alamos National Laboratory, Los Alamos,
NM 87545\\ $^3$Kavli Institute for Cosmological Physics and Department
of  Astronomy and  Astrophysics,  University of  Chicago, Chicago,  IL
60637}

\righthead{LENSING AND SUPERNOVAE}
\lefthead{SARKAR ET AL.}
\begin{abstract}
The  gravitational  magnification   and  demagnification  of  Type  Ia
supernovae  (SNe)  modify  their  positions  on  the  Hubble  diagram,
shifting    the    distance     estimates    from    the    underlying
luminosity-distance  relation.    This  can  introduce   a  systematic
uncertainty in the dark energy  equation of state (EOS) estimated from
SNe,  although  this  systematic  is  expected  to  average  away  for
sufficiently large data sets.   Using mock SN samples over the
redshift range $0 < z \leq 1.7$ we quantify the lensing bias.  We
find that the bias on the dark  energy EOS is less than half a percent
for  large  datasets  ($\gtrsim$   2,000  SNe).   However,  if  highly
magnified  events  (SNe  deviating   by  more  than  2.5$\sigma$)  are
systematically removed from the analysis, the bias increases to $\sim$
0.8\%.  Given that the EOS parameters measured from such a sample have
a  1$\sigma$  uncertainty of  10\%,  the  systematic  bias related  to
lensing in  SN data out to $z \sim 1.7$  can  be safely
ignored  in future  cosmological measurements.   
\end{abstract}

\section{Introduction}
\label{sec:introduction}

Since   the   discovery  of   the   accelerating   expansion  of   the
universe~\citep{Riess:98,  Perl:99, Knop:03,  Riess:04}, the  quest to
understand the physics responsible  for this acceleration has been one
of  the  major  challenges  of  cosmology.  At  present  the  dominant
explanation  entails  an additional  energy  density  to the  universe
called dark energy. The physics  of dark energy is generally described
in terms of its equation of  state (EOS), the ratio of its pressure to
density. In some  models this quantity can vary  with redshift.  While
there exist a variety of probes  to explore the nature of dark energy,
one of  the most compelling entails  the use of type  Ia supernovae to
map the  Hubble diagram, and thereby directly  determine the expansion
history of  the universe.  With increasing sample  sizes, SN distances
can potentially provide multiple independent estimates of the EOS when
binned   in   redshift   \citep{Huterer:05,Sullivan:07a,Sullivan:07b}.
Several present and  future SN surveys, such as  SNLS \citep{snls} and
the Joint  Dark Energy Mission  (JDEM), are aimed at  constraining the
value of the dark energy EOS to better than 10\%.

Although SNe  have been shown  to be good standardizable  candles, the
distance  estimate to  a given  SN  is degraded  due to  gravitational
lensing   of  its   flux  \citep{Frieman:97,  Wambsganss:97,
HolzWald:98}.  The  lensing becomes more prominent as  we observe SNe
out to higher  redshift, with the extra dispersion  induced by lensing
becoming  comparable   to  the  intrinsic   dispersion  (of  $\sim0.1$
magnitudes) at  $z \gtrsim 1.2$ \citep{Holz:05}.  In  addition to this
dispersion, which  leads to an  increase in the error  associated with
distance   estimate  to  each   individual  supernova,   lensing  also
correlates  distance estimates  of nearby  SNe on  the sky,  since the
lines-of-sight   pass   through   correlated  foreground   large-scale
structure  \citep{Cooray:05,Hui:06}.  Although this  correlation error
cannot be statistically eliminated by  increasing the number of SNe in
the  Hubble  diagram,  the  errors  can be  controlled  by  conducting
sufficiently wide-area ($> 5\mbox{  deg}^2$) searches for SNe (in lieu
of small-area pencil-beam surveys).

In addition  to the statistical  covariance of SN  distance estimates,
gravitational lensing also  introduces systematic uncertainties in the
Hubble  diagram  by  introducing  a  non-Gaussian  dispersion  in  the
observed  luminosities of  distant SNe.   Since lensing  conserves the
total  number  of  photons,  this  systematic bias  averages  away  if
sufficiently large  numbers of SNe  per redshift bin are  observed. In
this case the  average flux of the many  magnified and demagnified SNe
converges  on the unlensed  value \citep{Holz:05}.   Nonetheless, even
with thousands  of SNe  in the  total sample it  is possible  that the
averaging remains  insufficient, given  that one may  need to  bin the
Hubble diagram at very small redshift intervals to improve sensitivity
to the EOS. Furthermore, SNe at higher redshifts are more likely to be
significantly lensed.  If ``obvious''  outliers to the  Hubble diagram
are  removed from  the sample,  this introduces  an important  bias in
cosmological  parameter  determination,  and  can lead  to  systematic
errors in the determination of the dark energy EOS.

In this paper we  quantify the bias introduced in the estimation
of the  dark energy  EOS due  to weak lensing  of supernova  flux.  We
consider the  effects due  to the non-Gaussian  nature of  the lensing
magnification  distributions~\citep{Wang:02},  performing  Monte-Carlo
simulations by  creating mock  datasets for future  JDEM-like surveys.
The paper is organized as  follows: In $\S$2.1  we discuss our
parameterization   of   the  dark   energy   EOS,  $\S$2.2   discusses
gravitational  lensing, and $\S$3  is an  in-depth description  of our
methodology. We present our results in $\S$4.
\section{Background}
\label{sec:basics}

\subsection{Parameter Estimation Using SNe Ia}

For a SN with intrinsic luminosity $\mathcal L$,
the distance modulus is given by
\begin{equation}
m-M =-2.5 \log_{10} \left[ \frac{\mathcal L /4\pi {d_L}^2}{\mathcal L
/4\pi (10\,\mbox{pc})^2} \right] = 5 \log_{10} \left(
\frac{d_L}{\mbox{Mpc}}\right) + 25 ,
\label{eq:m-M}
\end{equation}
where, in  the framework of  FRW cosmologies, the  luminosity distance
$d_L$ is a  function of the cosmological parameters  and the redshift.
We   consider  the   two-parameter,  time-varying   dark   energy  EOS
\citep{ChevPol,Linder:03} adopted by the Dark Energy Task Force (DETF)
\citep{detf}.  We  take a flat universe described  by the cosmological
parameters  $\Theta =  \{  h, \Omega_m,  w_0,  w_a \}$,  with $h$  the
dimensionless  Hubble constant,  $\Omega_m$  the dimensionless  matter
density, and $w_0$ and $w_a$ the parameters describing the dark energy
EOS $w(z)=w_0+w_a(1-a)$. One can then express $d_L$ as
\begin{equation}
d_L(z) = (1+z) \int_0^z \frac{c dz'}{H(z')},
\label{eq:d_l_th}
\end{equation}
where $H(z)$ is the Hubble parameter at redshift $z$:
\begin{equation}
H(z) = H_0 \left[ \Omega_m(1+z)^3 + (1-\Omega_m) (1+z)^{3(1+w_0+w_a)}
          e^{-\frac{3w_az}{(1+z)}} \right]^{1/2}\,.
\end{equation}

\subsection{Weak Lensing of Supernova Flux}

Light  from a distant  SN passes  through the  intervening large-scale
structure of the universe.  This causes a modification of the observed
flux due to gravitational lensing:
\begin{equation}
{\mathcal F}^{\rm obs, lensed}(z, \hat{\bf n}) = \mu(z,\hat{\bf n})\,
             {\mathcal F}^{\rm obs, true},
\label{eq:mu} 
\end{equation}
where $\mu(z,  \hat{\bf n})$ is  the lensing induced  magnification at
redshift $z$ in the direction of the SN on the sky, $\hat{\bf n}$, and
${\mathcal  F}^{\rm obs,  true}$  is  the flux  that  would have  been
observed in  the absence  of lensing. The  magnification $\mu$  can be
either greater  than (magnified) or less than  (demagnified) one, with
$\mu=1$ corresponding to the (unlensed) pure FRW scenario.

This magnification (or demagnification)  of the observed flux leads to
an error in  the distance  modulus, which can be
expressed as
\begin{equation}
\left[\Delta (m-M)\right]_{\rm lensing} = - 2.5 \log_{10}
\left(\frac{{\mathcal F}^{\rm obs, lensed}}{{\mathcal F}^{\rm obs,
true}}\right) = - 2.5 \log_{10} \left(\mu \right)
\label{eq:delta(m-M)},
\end{equation} 
where we have written $\mu(z,\hat{\bf n})$ as $\mu$ for brevity.

Even in the absence of lensing the measured distance modulus to a Type
Ia SN suffers an intrinsic error, since the supernovae are not perfect
standard  candles. This  error is  typically  taken to  be a  Gaussian
distribution in either flux  or magnitude, with a redshift-independent
standard  deviation  of $\sigma_{\rm  int}$.  On  the  other hand,  as
lensing intrinsically  depends on  the optical depth,  which increases
with redshift, the scatter (or variance) due to lensing also increases
with redshift  \citep{Holz:05}. The probability  distribution function
(PDF) for lensing magnification, $P(\mu,z)$, of a background source at
redshift  $z$, depends  on both  the underlying  cosmology and  on the
nature of  the foreground structure responsible for  lensing.  We make
use  of  an  analytic  form  for $P(\mu)$  that  was  calibrated  with
numerical simulations \citep{Wang:02}, which  is valid for events that
are not strongly lensed ($\mu$ is less than a few).

\begin{figure}[!tb]
\begin{center}
\includegraphics[scale=0.5]{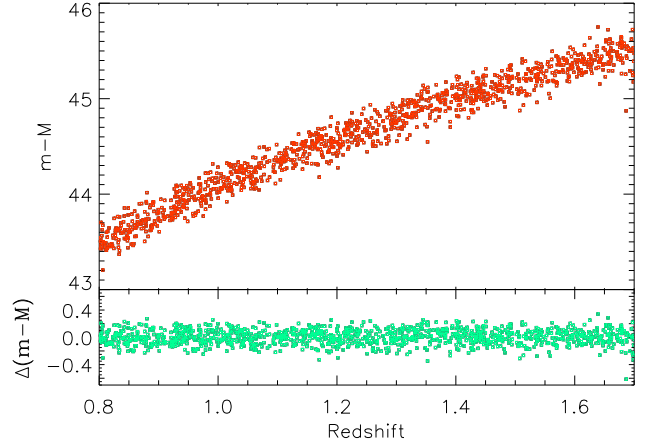}
\caption{{\it Top  panel:} The  Hubble diagram with  a linear
redshift scale showing the distance  modulus with redshift of a subset
of a  2000 SN mock dataset.  {\it Bottom panel:} The  residuals of the
data relative to the fiducial  model of a flat universe with $\Omega_m
= 0.3$, $w_0 = -1$, $w_a = 0$, and $h=0.732$. }
\label{fig:1}
\end{center}
\end{figure}

The generic  gravitational lensing PDF  peaks at a  demagnified value,
with  a  long tail  to  high  magnification \citep{Wang:02,
Sereno:02}. However,  since the total number of  photons is conserved
by lensing, all lensing  distributions preserve the mean: $\langle \mu
\rangle = \int\mu  P(\mu)\, d\mu=1$. To ensure that  this criterion is
met (to  an accuracy of  one part in  a million), we  re-normalize the
magnification PDF in a slightly different way than what was originally
suggested  in  \citet{Wang:02}.   $P(\mu)$  is  related,  at  a  given
redshift, to the probability distribution for the reduced convergence,
$\eta$,      through      $P(\mu)=P(\eta)/(2|\kappa_{min}|)$      (see
\citet{Wang:02};  Eq.   (6)),  where  $\kappa_{min}$  is  the  minimum
convergence.  The free parameter  in normalizing $P(\eta)$, and hence,
$P(\mu)$,  is   $\eta_{max}$,  the  maximum  value   for  the  reduced
convergence.   We determine  the  unique value  of $\eta_{max}$  which
yields both $\int P(\mu)\, d\mu=1$  and $\langle \mu \rangle= \int \mu
P(\mu)\, d\mu=1$  (to better than  $10^{-6}$).  Since the mode  is not
equal to  the mean, the distribution is  manifestly non-Gaussian.  The
majority  of  distant supernovae  are  slightly  demagnified, and  the
inferred  luminosity distances are  skewed to  higher values.   The SN
redshifts, on  the other hand, remain  unaffected, since gravitational
lensing   is   achromatic.     Although   lensing   magnification   is
insignificant at low  redshifts, at redshifts above one  both weak and
strong  lensing  become more  prominent.   For  small  SN samples  the
overall bias is towards  a larger cosmological acceleration (e.g., too
large  a  value of  $\Omega_\Lambda$  for  $\Lambda$CDM models).   The
lensing also  adds a systematic bias  to estimates of  the dark energy
EOS, shifting  to a more  negative value of  the EOS (e.g.,  less than
$-1.0$ for  a universe  with a cosmological  constant).  Along  with a
large fraction of demagnified SNe,  lensing produces a small number of
highly magnified sources.  If this high-magnification tail can be well
sampled,  the  average of  the  lensing  distribution  is expected  to
converge  on   the  true  mean,   and  the  bias  can   be  eliminated
\citep{Wang:00,Holz:05}.

\section{Methodology}
\label{sec:analysis}

\begin{figure*}[]
\centerline{
\includegraphics[scale=0.5]{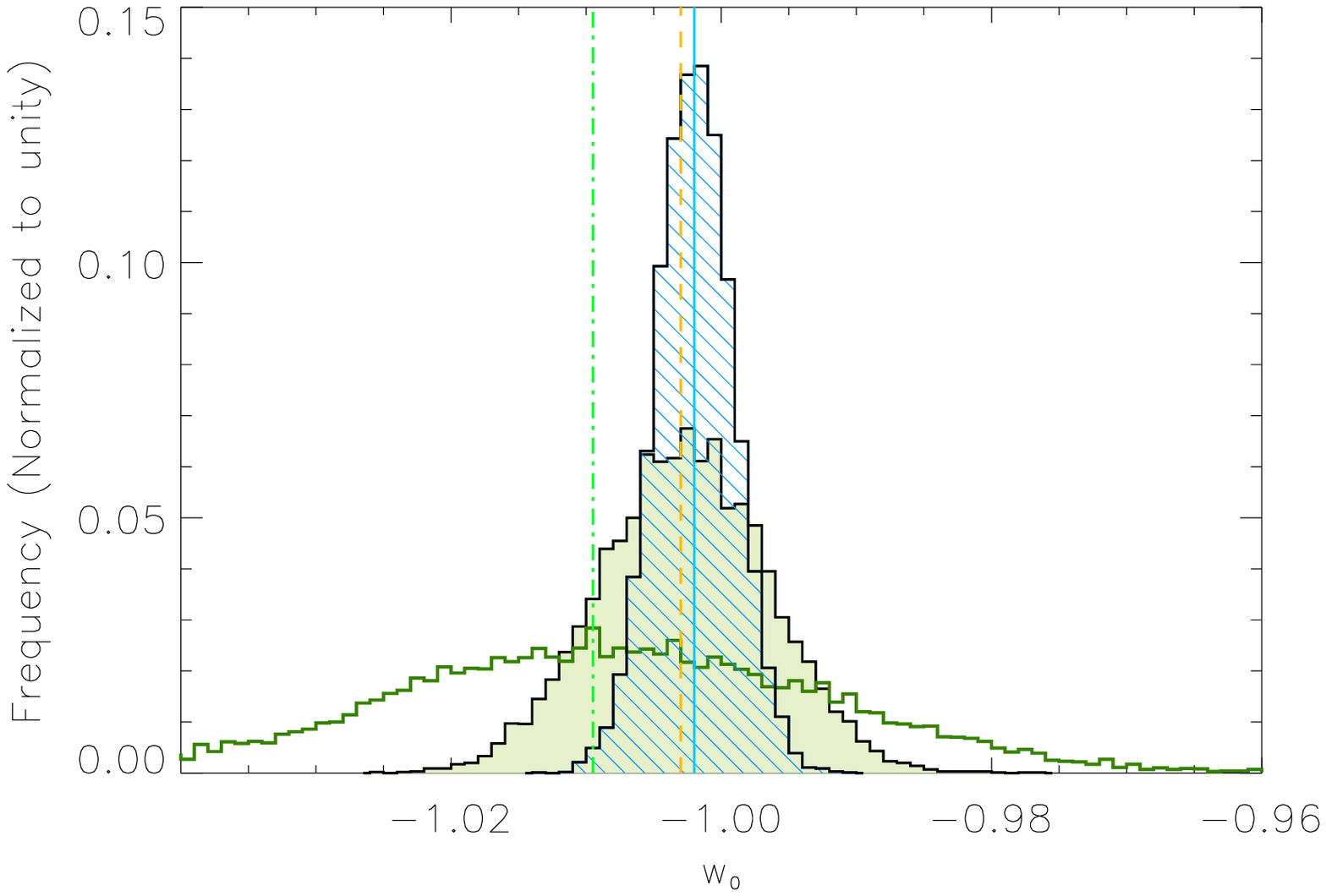}
\includegraphics[scale=0.5]{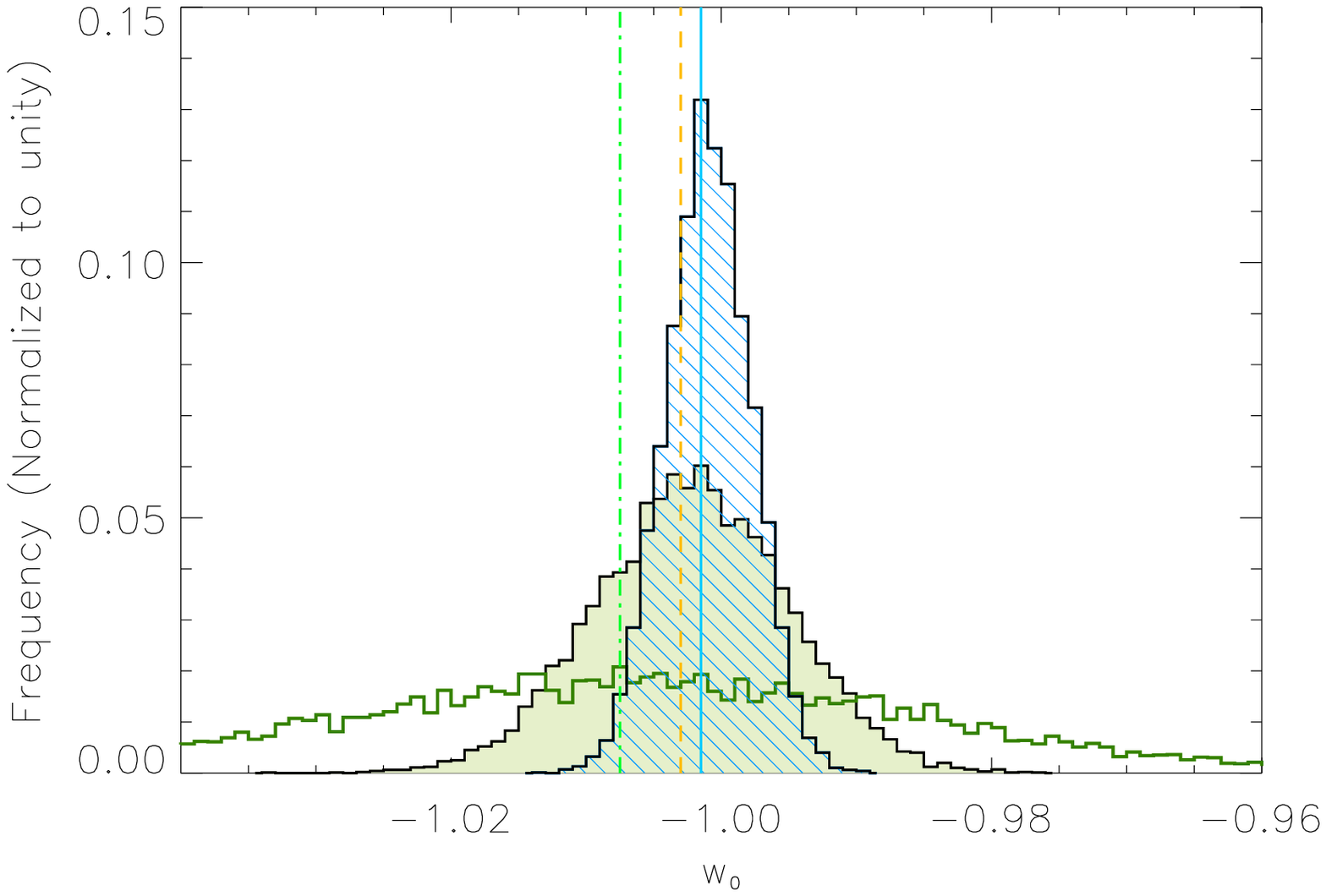}
}

\caption{{\it Left panel:}  Histograms showing the distribution (after
10,000  realizations)  of the  values  obtained  for  $w_0$ using  the
magnitude analysis  (\S~3.1), after marginalizing over  $w_a$ and $h$,
for three different sample sizes.  The un-filled histogram depicts the
case of the 300 SN sample, the shaded histogram shows the distribution
for a  sample size of 2000,  and the hatched histogram  shows the case
for  10,000 SN  sample.  The  vertical  lines at  -1.0095 (dashed  and
dotted), -1.003  (dashed), and -1.002  (solid) show the  average $w_0$
values for  the 300  SN, 2000 SN,  and 10,000 SN  cases, respectively.
The width of the distributions is due primarily to the intrinsic noise
of the  SNe, while the shifted mode  is due to finite  sampling of the
gravitational lensing PDF.  {\it  Right panel:} Histograms showing the
distribution (after  10,000 realizations)  of the values  obtained for
$w_0$ using the flux-averaging technique (\S~3.2), after marginalizing
over $w_a$ and  $h$, for three different sample  sizes.  The un-filled
histogram depicts the case of  the 300 SN sample, the shaded histogram
shows  the distribution for  a sample  size of  2000, and  the hatched
histogram shows the case for  10,000 SN sample.  The vertical lines at
-1.007 (dashed and dotted),  -1.0025 (dashed), and -1.001 (solid) show
the  $w_0$  values that  were  obtained  after  averaging over  10,000
realizations   for   the  300,   2000,   and   10,000  sample   sizes,
respectively.}
\label{fig:2}
\end{figure*}

We generate mock  SN samples, and quantify the  bias in the estimation
of the  dark energy EOS due  to gravitational lensing. We  pick as our
fiducial cosmology a flat universe  with $\Omega_m = 0.3$, $w_0 = -1$,
$w_a = 0$, and $h=0.732$ with a Gaussian prior of $\sigma(h) = 0.032$.
We consider a  range of possible upcoming surveys,  and Monte-Carlo SN
samples of  varying sizes distributed uniformly in  the redshift range
$0.1<z<1.7$.  If the SN intrinsic errors are Gaussian in flux, then it
makes sense to analyze the data  via flux averaging. In this case both
the  intrinsic errors  and the  lensing errors  will average  away for
sufficiently large  SN samples. However,  if the intrinsic  errors are
Gaussian  in  magnitude,  then   a  magnitude  analysis  may  be  more
appropriate, although this  will lead to a bias  due to lensing (which
is only bias-free for flux  analysis). In what follows we perform both
analyses.

\subsection{Magnitude Analysis}
\label{sec:mag}

For each SN at a redshift $z$  we draw a value of $\mu$ at random from
the   re-normalized    lensing   PDF,   $P(\mu,z)$,    obtained   from
\citet{Wang:02},   and  use   Eq.~(\ref{eq:delta(m-M)})   to  evaluate
$\left[\Delta  (m-M)\right]_{\rm lensing}$.   To  model the  intrinsic
scatter we  draw a  number at random  from a Gaussian  distribution of
zero mean  and standard deviation given by  $\sigma_{\rm int}=0.1$ mag
(\citet{snls} find  a scatter at the  level of 0.13  mag).  We combine
the intrinsic and lensing  noise with the ``true'' underlying distance
modulus to get the observed distance modulus:
\begin{equation}
(m-M)^{\rm    data}(z)   =    (m-M)^{\rm   fid}(z)    +   \left[\Delta
(m-M)\right]_{\rm int} + \left[\Delta (m-M)\right]_{\rm lensing} \, .
\end{equation}
By repeating the  above procedure for each of the SNe  in a sample, we
generate a mock Hubble diagram.  We create mock surveys with different
numbers of  SNe: $N= 300$,  2000, \& 10,000;  and then for  each fixed
number of SNe we generate  at least 10,000 independent mock samples to
properly sample the  distributions.  In Figure~\ref{fig:1} we
show an example of  one such dataset  with a SNe sample size of
2000. The upper  panel shows  the Hubble diagram  and the  lower panel
shows the residuals of the distance moduli relative to that of the 
fiducial cosmological model.

For each mock Hubble diagram  we compute the likelihood of a parameter
set ${\bf p}$ by evaluating the $\chi^2$-statistic:
\begin{equation}
\chi^2({\bf p}) = \sum_{i=1}^{N} \frac{\left[ (m-M)^{\rm data}(z_i)-
(m-M)^{\rm fid}(z_i)\right]^2}{\sigma^2(z_i)} \, ,
\label{eq:chi}
\end{equation} 
where $\sigma(z_i)$ is  the error bar for the  distance modulus of the
$i$th supernova, taken to be 0.1 mag throughout. The projected bias in
the  estimation of  $w_0$ can  now be  computed by  marginalizing over
$w_a$ and $h$ (keeping $\Omega_m$ fixed).

\subsection{Flux-Averaging Analysis}
\label{sec:flux}

\citet{Wang:00}  argues that  averaging the  flux, as  opposed  to the
magnitudes, of observed SNe  naturally removes the lensing bias, since
the mean  of the lensing distributions  are equal to one  in flux, but
not  in  magnitude.   We thus  also  analyze  the  mock data  sets  in
flux. The flux averaging is done such that for each supernova we first
use Eq.~(\ref{eq:d_l_th}) to calculate the value of $d_L(z)$, and then
evaluate the fiducial flux using
\begin{equation}
{\mathcal   F}^{\rm  fid}   (z)=   \frac{\mathcal  L}{4\pi   [d_L^{\rm
fid}(z)]^2},
\label{eq:flux}
\end{equation} 
where we can  take any {\it a priori} fixed value  of $\mathcal L$. We
take  the lensing  PDF,  $P(\mu)$,  and convolve  it  with a  Gaussian
distribution having zero mean  and dispersion 0.1 (this corresponds to
$\sim$ 5\% error  in distance estimates).  We then  randomly draw from
the  convolved distribution,  and multiply  this number  by ${\mathcal
F}^{\rm fid}(z)$  to get the observed  flux for a  given supernova. We
repeat this  process for  each of the  $N$ SNe  to obtain a  mock data
set. We then  follow the flux averaging recipe  of \citet{Wang:04} and
perform the  likelihood analysis  to quantify the  bias. As  before we
perform the above  simulation a large ($\sim$ 10,000)  number of times
to get a distribution for the bias in parameter estimation.

We further assume that our SNe  samples are complete in the sense
that there is no Malmquist bias  or any bias due to detection effects over the
redshift range considered. This is a reasonable assumption since we consider
SNe out to a redshift of 1.7 and such a complete catalog is expected from the
near-infrared imaging capabilities of the JDEM program.

\section{Results and Discussion}
\label{sec:discussion}

We first present  our results for the magnitude  case, as described in
\S\ref{sec:mag}.    The  left   panel   of  Figure~\ref{fig:2}   shows
histograms  of  the  best-fit  values  of $w_0$  from  the  likelihood
analysis,  after   marginalizing  over  $w_a$  and   $h$.   The  empty
histogram, which  peaks at -1.009  (marked with a  vertical dot-dashed
line),  is  for  the  model  with  300  SNe.   The  shaded  histogram,
representing  the 2,000  SN  case, peaks  at  -1.003 (vertical  dashed
line), while the hatched histogram representing the 10,000 SN case has
its peak  at -1.002 (vertical  solid line).  These  distributions have
1$\sigma$  widths  of 0.016,  0.006,  and  0.003, respectively.   This
scatter is primarily due  to the intrinsic uncertainty associated with
absolute calibration,  and is not  dominated by lensing.   Without the
inclusion of  lensing, however, the distributions peak  at exactly -1,
and show no bias.  The shifted mode  gives us a rough idea of the bias
to be expected, on average, due  to lensing.  We find that 68\% of the
time a random sample of 300 SNe will have an estimated value for $w_0$
within  3\% of  its fiducial  value, and  this drops  to 0.5\%  when a
sample size of 10,000 SNe is considered.

The right panel of  Figure~\ref{fig:2} shows the same distributions as
the  left panel,  but  this time  using  the flux-averaging  technique
instead of averaging over  magnitudes.  The empty, shaded, and hatched
histograms peak  at -1.007, -1.003, and  -1.001, respectively, showing
the mean  bias for  the 300,  2,000 and 10,000  SN cases  (marked with
dot-dashed, dashed,  and solid vertical  lines).  With flux-averaging,
we expect that 68\% of the time  a random sample of 300 SNe will yield
a value of $w_0$ within 2.5\%  of the fiducial value, and within 0.5\%
for a sample of 10,000 SNe.

The  1$\sigma$ parameter  uncertainty on  $w_0$ ranges  from  the 20\%
level (for  300 SNe) to less  than 5\% (for 10,000  SNe), dwarfing the
bias due  to lensing.   Thus, we need  not be concerned  about lensing
degradation  of  dark  energy  parameter estimation  for  future  {\it
JDEM}-like surveys.  We note, however,  that our estimated bias on the
EOS is larger than the lensing  bias of $w<0.001$ quoted in Table 7 of
\citet{Wood:07}.   This is  not  surprising, given  their  use of  the
simple Gaussian  approximation to lensing  from \citet{Holz:05}, which
is less effective for low statistics. Nonetheless, we agree with their
conclusion that lensing  is negligible.  A similar conclusion
was also reached by \citet{Martel:07} who used  a compilation of
230 Type Ia  SNe \citep{Tonry:03} in the redshift range $0  < z < 1.8$
to show  that the lensing errors  are small compared  to the intrinsic
SNe errors.

\begin{figure}[!tb]
\begin{center}
\includegraphics[scale=0.5]{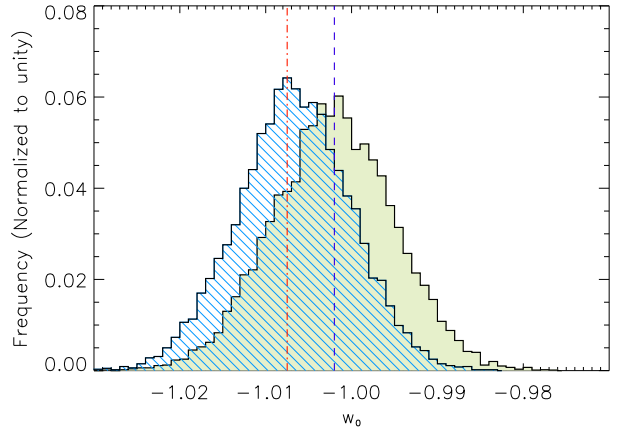}
\caption{Histograms  showing the distribution  of the  values obtained
for $w_0$ after marginalizing over $w_a$ and $h$ for the 2,000 SN case
(using  the flux-averaging technique).   The shaded  histogram assumes
that   the  full   sample  of   2,000  SNe   is  used   for  parameter
estimation. The  hatched histogram shows the shift  when outliers (SNe
that are shifted  above or below the Hubble diagram  by more than 25\%
on  either  side)  are  removed  from  the sample.  The  bias  in  the
distribution is due to  the removal of highly-magnified lensing events
from the sample.}
\label{fig:3}
\end{center}
\end{figure}

We now discuss the bias which arises if anomalous SNe are removed from
the  sample.   Gravitational lensing  causes  some  SNe  to be  highly
magnified, and  it is conceivable that these  ``obvious'' outliers are
subsequently removed from  the analysis. In this case  the mean of the
sample will be  shifted away from the true  underlying Hubble diagram,
and a bias will be  introduced in the best-fit parameters. To quantify
this  effect, we  remove  SNe  which deviate  from  the expected  mean
luminosity-distance relation  in the Hubble diagram by  more than 25\%
(corresponding roughly to a $2.5\sigma$  outlier). The SN scatter is a
result of the convolution of  the intrinsic error (Gaussian in flux of
width 0.1)  and the  lensing PDF,  and the outlier  cutoff leads  to a
removal of  $\sim$ 50 SNe  out of the  2,000 SNe.  These  outliers are
preferentially  magnified,  due to  the  strong  lensing  tail of  the
magnification distributions.   The demagnification tail is  cut off by
the empty-beam  lensing limit, and  therefore isn't as  prominent. The
hatched  histogram in Figure~\ref{fig:3}  shows the  distribution when
events with convolved error greater than 2.5$\sigma$ are removed.  The
vertical dot-dashed line  at -1.0075 shows the average  value of $w_0$
obtained in  this case, representing a  bias in the  estimate of $w_0$
roughly three times  larger than when the full  2,000 SNe are analyzed
(shown by shaded histogram).  This bias is a result of cutting off the
high  magnification tail of  the distribution,  and thus  shifting the
data towards a net dimming of observed SNe, leading to a more negative
value of $w_0$.

We also  apply a cutoff at  $3\sigma$, in addition  to the $2.5\sigma$
discussed above. This results in a removal of $\sim20$ SNe on average,
for each  2,000 SN  sample, and  leads to a  bias of  $\sim0.6$\%. Any
arbitrary cut  on the  (non-Gaussian) convolved (lensing  + intrinsic)
sample leads  to a  net bias  in the distance  relation, and  even for
large outliers  and large SN  samples, this can lead  to percent-level
bias in the best-fit values for $w_0$.

To  summarize, we  have quantified  the effect  of  weak gravitational
lensing on  the estimation of dark  energy EOS from  type Ia supernova
observations.  With  generated mock  samples of 2,000  SNe distributed
uniformly in redshift up to  z$\sim$1.7 (as expected in future surveys
like {\it JDEM}), we have  shown that the bias in parameter estimation
due to lensing is less than  1$\%$ (which is well within the 1$\sigma$
uncertainty expected  for these missions). Analyzing the  data in flux
or magnitude  does not alter  this result.  If lensed  supernovae that
are highly magnified (such that  the convolved error is more than 25\%
from the  underlying Hubble  diagram) are systematically  removed from
the sample,  we find  that the  bias increases by  a factor  of almost
three. Thus, so long as all  observed SNe are used  in the Hubble
diagram, including  ones that  are highly magnified,  the bias  due to
lensing in the  estimate of the dark energy  EOS will be significantly
less  than the  1$\sigma$  uncertainty.  Even  for  a post-{\it  JDEM}
program with 10,000 SNe, lensing bias can be safely ignored.

\acknowledgements  We  thank  Yun  Wang for  useful  discussions.   AC
acknowledges support  from NSF CAREER AST-0645427.  DEH acknowledges a
Richard  P. Feynman  Fellowship from  Los Alamos  National Laboratory.
DS, \ AC, \ and DEH\ are partially supported by the DOE at LANL and UC
Irvine  through IGPP  Grant  Astro-1603-07.  AA  acknowledges a  McCue
Fellowship at UC Irvine.


\begin{thebibliography}{}

\bibitem[Aldering et al.(2004)]{snap}
 Aldering, G. et al.\ 2004, PASP, submitted (astro-ph/0405232)

\bibitem[Astier et al.(2006)]{snls}
Astier, P. et al., \ 2006, A \& A, 447, 31

\bibitem[Chevalier \& Polarski (2001)]{ChevPol}
  M.~Chevallier and D.~Polarski,
  Int. J. Mod. Phys. D {\bf 10}, 213 (2001).

\bibitem[Cooray, Huterer, \& Holz(2006)]{Cooray:05}
Cooray, A., Huterer, D., Holz, D. 2006, PRL, 96, 021301

\bibitem[Frieman(1997)]{Frieman:97}
Frieman, J. A.\ 1997, Comments Astrophys., 18, 323

\bibitem[Albrecht et al. (2006)]{detf}
Albrecht, A. et al.\ 2006, arXiv:astro-ph/0609591

\bibitem[Holz \& Linder(2005)]{Holz:05}
Holz, D. E. \& Linder, E. V. \ 2005, \apj, 631, 678

\bibitem[Holz \& Wald(1998)]{HolzWald:98}
Holz, D. E. \& Wald, R. M. 1998, PRD, 58, 063501

\bibitem[Hui \& Greene(2006)]{Hui:06}
Hui, L. \& Greene, P. B. \ 2006, PRD, 73, 123526

\bibitem[Huterer \& Cooray(2005)]{Huterer:05}
Huterer, D. \& Cooray, A.\ 2005, PRD, 71, 023506

\bibitem[Knop et al.(2003)]{Knop:03}
Knop, R. A. et al.\ 2003, \apj, 598, 102

\bibitem[Linder (2003)]{Linder:03}
Linder, E. \ 2003, PRL, 90, 091301


\bibitem[Martel \& Premadi (2007)]{Martel:07}
Martel, H. \& Premadi, P.  \ 2007, arXiv:0710.5452

\bibitem[Perlmutter et al.(1999)]{Perl:99}
Perlmutter, S., et al. \ 1999, \apj, 517, 565

\bibitem[Riess et al.(1998)]{Riess:98}
Riess, A. G. et al.\ 1998, \aj, 116, 1009

\bibitem[Riess et al.(2004)]{Riess:04}
Riess, A. G. et al.\ 2004, \apj, 607, 665

\bibitem[Sereno, Piedipalumbo, \& Sazhin (2002)]{Sereno:02}
Sereno, M., Piedipalumbo, E., Sazhin, M.V. \ 2002, MNRAS, 335, 1061

\bibitem[Sullivan, Cooray, \& Holz(2007)]{Sullivan:07a} 
Sullivan, S., Cooray,   A., \& Holz, D. E. \ 2007, JCAP, 09, 004

\bibitem[Sullivan  et al.(2007)]{Sullivan:07b} Sarkar, D., Sullivan,  S., Joudaki,  S., Amblard, A.,  Holz, D.  E.,  \& 
Cooray, A.  \ 2007, arXiv:astro-ph/0709.1150

\bibitem[Tonry et al. (2003)]{Tonry:03}
Tonry, J. L. et al. \ 2003, \apj, 594, 1 

\bibitem[Wambsganss et al. (1997)]{Wambsganss:97}
Wambsganss, J., Cen, R., Xu, G., \& Ostriker, J. P. \ 1997,\apj, 475, L81

\bibitem[Wang(2000)]{Wang:00}
Wang, Y.\ 2000, \apj, 536, 531

\bibitem[Wang, Holz, \& Munshi(2002)]{Wang:02}
Wang, Y., Holz, D. E., Munshi, D.\ 2002, \apj, 572, L15

\bibitem[Wang \& Mukherjee(2004)]{Wang:04}
Wang, Y. \& Mukherjee, P.\ 2004, \apj, 606, 654

\bibitem[Wood-Vasey et al. (2007)]{Wood:07}
 W.~M.~Wood-Vasey {\it et al.},
  arXiv:astro-ph/0701041.


\end{thebibliography}
\end{document}